\documentclass[aps,pre,superscriptaddress,twocolumn,showpacs,showkeys]{revtex4}
\usepackage{graphicx}
\usepackage{psfrag}
\usepackage{amsmath}

\begin{document}

\title{Curie temperature for an Ising model on Archimedean lattices}

\author{K. Malarz}
\homepage{http://home.agh.edu.pl/malarz/}
\email{malarz@agh.edu.pl}
\affiliation{
Faculty of Physics and Applied Computer Science,
AGH University of Science and Technology,\\
al. Mickiewicza 30, PL-30059 Krak\'ow, Poland.
}
\author{M. Zborek}
\affiliation{
Faculty of Physics and Applied Computer Science,
AGH University of Science and Technology,\\
al. Mickiewicza 30, PL-30059 Krak\'ow, Poland.
}
\author{B. Wr\'obel}
\affiliation{
Faculty of Physics and Applied Computer Science,
AGH University of Science and Technology,\\
al. Mickiewicza 30, PL-30059 Krak\'ow, Poland.
}

\date{\today}

\begin{abstract}
Critical temperatures for the ferro-paramagnetic transition in the Ising model are evaluated for five Archimedean lattices, basing on Monte Carlo simulations.
The obtained Curie temperatures are $1.25$, $1.40$, $1.45$, $2.15$ and $2.80$ $[J/k_B]$ for $(3,12^2)$, $(4,6,12)$, $(4,8^2)$, $(3,4,6,4)$ and $(3^4,6)$ lattices, respectively.
\end{abstract}



\maketitle

\section{Introduction}
\label{sec-intro}

Beauty of the Ising model (IM) \cite{ising,books} manifests in its simplicity.
The system considered is a network of $N$ interacting spins $S_i=\pm 1$ which energy is
\begin{equation}
E \equiv -J\sum_{(i,j)} S_i S_j,
\label{Ham}
\end{equation}
where $J$ is an exchange integral.
We assume homogeneous short range spin interactions, i.e.
the summation in Eq. \eqref{Ham} goes over pairs $(i,j)$ of nearest neighbours.
Positive sign of $J>0$ gives ferromagnetic interaction among spins.
The minimisation of energy \eqref{Ham} for temperature $T=0$ produces spin dynamics which may be described by deterministic cellular automaton with rule
\begin{equation}
S_i(t+1)=\text{sign}\left(J \sum_j S_j(t)\right),
\label{ca-rule}
\end{equation}
where $t$ denotes discrete time and summation goes over nearest neighbours of $i$-th spin.

For finite temperature $T>0$ the deterministic rule \eqref{ca-rule} is replaced by a probabilistic cellular automaton with spin update rule $S_i(t)\to S_i(t+1)$ described by Glauber \cite{glauber} or Metropolis \cite{metropolis} dynamics.
Then the phase transition may be observed: below critical temperature $T<T_C$ spontaneous magnetisation $m\equiv\sum_{i=1}^N S_i/N\ne 0$ is observed while $m=0$ for $T>T_C$.

The IM was already investigated
\begin{itemize}
\item in many ways, including Monte Carlo simulation \cite{mcs}, series expansion \cite{series-exp,adler}, mean-field approach \cite{mfa} or partition function technique \cite{partition},
\item and for many systems, for example: antiferromagnets \cite{tadic,gosia,af}, frustrated \cite{tadic,gosia,frustrated}, disordered \cite{disord} or diluted \cite{diluted} networks on complex \cite{tadic,complex} or shuffled lattices \cite{crazy}, etc. \cite{new}.
\end{itemize}

In this paper the critical temperature $T_C$ is estimated for five two-dimensional lattices, basing on $\langle m(T)\rangle$ dependence, where $\langle\cdots\rangle$ denotes the time average. 
The Archimedean lattices are vertex transitive graphs that can be embedded in a plane such that every face is a regular polygon.
Kepler showed that there is exactly eleven such graphs \cite{kepler}.
The names of the lattices are given according to the sizes of faces incident to a given vertex.
The face sizes are listed in order, starting with a face such that
the list is the smallest possible in the lexicographical order.
In this way, the square lattice gets the name $(4, 4, 4, 4)$, abbreviated to $(4^4)$, honeycomb is called $(6^3)$ and Kagom\'e is $(3, 6, 3, 6)$. 
Some results concerning IM on AL were already presented in Refs. \cite{archi,lin,jpa7,dixon,kramers,onsager}, however in a literature known to us the Curie temperatures of several AL are still missing.

Critical properties of these lattices were investigated in terms of site percolation \cite{percol-ds} in Ref. \cite{percolation}.
Topologies of all eleven AL are given there as well.

\section{Results of the simulations}
\label{sec-res}

\begin{figure}
\psfrag{<m>}{$\langle m\rangle$}
\psfrag{TJkB}{$T\, [J/k_B]$}
\includegraphics[width=0.45\textwidth]{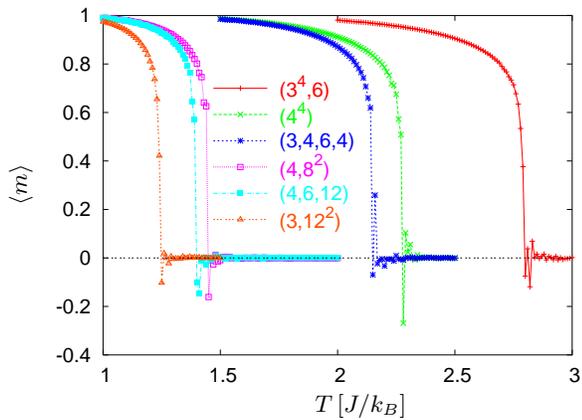}
\caption{Dependence of the average magnetisation $\langle m\rangle$ on temperature $T$ expressed in $[J/k_B]$ units for $(3^4,6)$, $(3,4,6,4)$, $(4,6,12)$, $(4,8^2)$ and $(4^4)$ AL.
The simulations are carried out for $N\approx 6\cdot 10^4$ spins during $N_{\text{iter}}=2\cdot 10^5$ [MCS].
The magnetisation $\langle m\rangle$ is averaged over the last $10^5$ [MCS].}
\label{fig-mvsT}
\end{figure}

\begin{table}
\caption{AL and associated critical temperatures $T_C$.}
\label{tab-tc}
\begin{ruledtabular}
\begin{tabular}{llll}
$z$ & lattice    & $T_C$ $[J/k_B]$ & Ref.\\
\hline
\hline
3 & $(3,12^2)$   & ${\bf 1.25}$ & --- \\ 
  & $(4,6,12)$   & ${\bf 1.40}$ & --- \\ 
  & $(4,8^2)$    & ${\bf 1.45}$ & --- \\ 
  & $(6^3)$      & $1.52$ & \cite{dixon} \\ 
\hline
4 & $(3,4,6,4)$  & ${\bf 2.15}$ & --- \\
  & $(4^4)$      & $2/\text{arcsinh}\,1\approx 2.27$&\cite{kramers,onsager}\\
  & $(3,6,3,6)$  & $2.27$       & \cite{adler}\\ 
\hline
5 & $(3^4,6)$    & ${\bf 2.80}$ & --- \\
  & $(3^3,4^2)$  & $2/\ln 2\approx 2.89$ & \cite{jpa7} \\
  & $(3^2,4,3,4)$& $2.93$       & \cite{jpa7} \\
\hline
6 & $(3^6)$      & $3.64$       & \cite{dixon} \\ 
\end{tabular}
\end{ruledtabular}
\end{table}

We evaluate the Curie temperature $T_C$ basing on the termal dependence of magnetisation $\langle m\rangle$.
Investigated systems contain about $N\approx 6\cdot 10^4$ spins, which decorate nodes of $(3^4,6)$, $(3,4,6,4)$, $(4,6,12)$, $(4,8^2)$ AL. 
The Glauber dynamics \cite{glauber} is applied and the simulation takes $N_{\text{iter}}=2\cdot 10^5$ Monte Carlo steps (MCS).
One MCS is completed when all $N$ spins are investigated (spin-by-spin in a type-writer order), wheter they should flip or not.
The time average is performed over the last $10^5$ MCS for an evaluation of $\langle m\rangle$.
The results are presented in Fig. \ref{fig-mvsT}.
Temperature for which spontaneous magnetisation $\langle m\rangle$ vanishes is accepted to be an estimation of $T_C$.
These estimations are shown in Tab. \ref{tab-tc}.

\section{Conclusions}
In this paper the Curie temperatures for IM on all AL are collected.
Among them, $T_C$ for $(3,12^2)$, $(4,6,12)$, $(4,8^2)$, $(3,4,6,4)$ and $(3^4,6)$ AL are evaluated for the first time with the Monte Carlo simulation.

\begin{figure}
\psfrag{<m>}{$\langle m\rangle$}
\psfrag{TTC}{$\kappa=T/T_C$}
\includegraphics[width=0.45\textwidth]{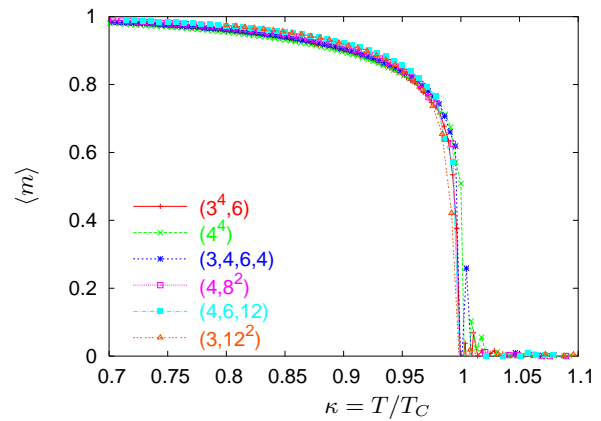}
\caption{Dependence of the average magnetisation $\langle m\rangle$ on normalized dimensionlees temperature $T/T_C$ for $(3^4,6)$, $(3,4,6,4)$, $(4,6,12)$, $(4,8^2)$ and $(4^4)$ AL.
The same data as in Fig. \ref{fig-mvsT}.}
\label{fig-norm}
\end{figure}

For all investigated AL, the shape of $m(T/T_C)$ curve (see Fig. \ref{fig-norm}) is roughly the same as for the square lattice.
In the latter case, an analytical expression \cite{yang} is known 
\[
|m(\kappa)|=\sqrt[8]{\dfrac{\cosh^2(2/\kappa)}{\sinh^4(2/\kappa)}[\sinh^2(2/\kappa)-1]},
\]
where $\kappa\equiv T/T_C$.

In contrast to Galam--Mauger \cite{Tc_alld} semi-exact formula for $T_C$ dependence on system dimensionality $d$ and lattice coordination number $z$, we show that critical temperature for IM differ slightly for several AL (where $d=2$) with the same values of $z$.
Similarly to the percolation phenomena \cite{percol-ds}, also for IM the dimensionality $d$ and the coordination number $z$ are not sufficient \cite{pcdz,marck,km-sg} for determining the critical point $T_C$.

\begin{acknowledgments}
K.M. is grateful to Krzysztof Ku{\l}akowski for many valuable and fruitful discussions.
Calculations were carried out in ACK-CYFRONET-AGH.
The machine time on HP Integrity Superdome is financed by the Polish Ministry of Science and Information Technology under Grant No. MNiI/HP\_I\_SD/AGH/047/2004.
\end{acknowledgments}


\end{document}